ORDER, DISORDER, AND PHASE TRANSITION
IN CONDENSED SYSTEMS

# Localization Theory in Zero Dimension and the Structure of the Diffusion Poles

**I. M. Suslov**

*Kapitza Institute for Physical Problems, Russian Academy of Sciences, ul. Kosygina 2, Moscow, 119334 Russia*
*e-mail: suslov@kapitza.ras.ru*
Received June 9, 2007

**Abstract**—The $1/[-i\omega + D(\omega, q)q^2]$ diffusion pole in the localized phase transfers to the $1/\omega$ Berezinskii–Gorkov singularity, which can be analyzed by the instanton method {M. V. Sadovskiĭ, Zh. Éksp. Teor. Fiz. **83**, 1418 (1982) [Sov. Phys. JETP **56**, 816 (1982)] and J. L. Cardy, J. Phys. C **11**, L321 (1978)}. Straightforward use of this approach leads to contradictions, which do not disappear even if the problem is extremely simplified by taking zero-dimensional limit. On the contrary, they are extremely sharpened in this case and become paradoxes. The main paradox is specified by the following statements: (i) the $1/\omega$ singularity is determined by high orders of perturbation theory, (ii) the high-order behaviors for two quantities $\Phi^{RA}$ and $U^{RA}$ are the same, and (iii) $\Phi^{RA}$ has the $1/\omega$ singularity, whereas $\Phi^{RA}$ does not have it. Solution to the paradox indicates that the instanton method makes it possible to obtain only the $1/(\omega + 2i\gamma)$ singularity, where the parameter $\gamma$ remains indefinite and must be determined from additional conditions. This conceptually confirms the necessity of the self-consistent treatment for the diffusion coefficient that is used in the Vollhardt–Wölfle type theories.

## 1. INTRODUCTION

The modern formulation of the Anderson transition problem is as follows [1]. The quantity

$$\Phi^{RA}(x_1, x_2, x_3, x_4) = \langle G^R_{E+\omega}(x_1, x_2) G^A_E(x_3, x_4) \rangle, \quad (1)$$

which contains complete information on the kinetic properties, is introduced, where $G^R$ and $G^A$ are the retarded and advanced Green's functions for the electron in a random potential, respectively. The Fourier transform $\Phi^{RA}(q)$ of this quantity at the coinciding arguments $x_1 = x_4$ and $x_2 = x_3$ has the diffusion pole (see, e.g., [2, 3])

$$\Phi^{RA}(q) = \frac{2\pi\nu(E)}{-i\omega + D(\omega, q)q^2} + \Phi_{\text{reg}}(q), \quad (2)$$

where $D(\omega, q)$ is the observed diffusion coefficient and $\nu(E)$ is the density of states near the energy $E$. The diffusion pole in the localized phase transfers to the $1/\omega$ Berezinskii–Gorkov singularity

$$\Phi^{RA}(q) = \frac{2\pi\nu(E)}{-i\omega} A(q) + \Phi_{\text{reg}}(q). \quad (3)$$

Coordination of Eqs. (2) and (3) provides the conclusion that $D(\omega, q) = -i\omega d(q)$ in the localized phase, so that $D(0, q) \equiv 0$. Therefore, the Anderson transition does not reduce to the disappearance of $D(0, 0)$ and is of deeper meaning [7]: the function $D(0, q)$ is identically equal to zero at the transition point [1, 3]. The complicated rearrangement of $D(\omega, q)$ near the transition point is of main interest for the theory.

A qualitative scenario of the transition was proposed by Vollhardt and Wölfle [2] and corresponds to the following simple estimate. The irreducible vertex $U^{RA}_{\mathbf{kk}'}(\mathbf{q})$ entering into the Bethe–Salpeter equation has the diffusion pole in the limit $\mathbf{k} + \mathbf{k}' \longrightarrow 0$:

$$U^{RA}_{\mathbf{kk}'}(\mathbf{q}) = U^{\text{reg}}_{\mathbf{kk}'}(\mathbf{q}) + \frac{F(\mathbf{k}, \mathbf{k}', \mathbf{q})}{-i\omega + D(\omega, \mathbf{k}+\mathbf{k}')(\mathbf{k}+\mathbf{k}')^2} \quad (4)$$

and serves as the "transition probability" in the quantum kinetic equation. The use of the $\tau$ approximation, $D \propto \tau \propto \langle U \rangle^{-1}$ (where $\langle \ldots \rangle$ is averaging over momenta), leads to the equation

$$D \sim \text{const} \left( U_0 + F_0 \int \frac{d^d q}{-i\omega + D(\omega, q) q^2} \right)^{-1} \quad (5)$$

When the spatial dispersion of $D(\omega, q)$ is disregarded, this equation describes the transition between the $D \sim i\omega$ and $D = \text{const}(\omega)$ regimes [2]. A more refined variant of the theory, where the spatial dispersion of $D(\omega, q)$ is taken into account,[1] was developed in our work [3]. This variant has good prospects to be exact, but in the current status, it is in incomplete agreement with the numerical researches [8] (see discussion in [9]). For this reason, the direct microscopic verification of the phenomenological scheme proposed in [3] is of current interest.

This study is devoted to the problem of the structure of the diffusion pole in the region of the fluctuation tail of the density of states, i.e., in the energy region corresponding to the band gap of the initial material. According to [5, 10], the diffusion pole in this case is of purely nonperturbative origin and can be found by the instanton method.[1] The idea of such an approach is undoubtedly correct. However, a number of problems concerning the calculation procedure and results arise.

(i) The saddle-point approximation used in [5, 10] makes it possible to obtain the dependence $\Phi^{RA} \sim 1/\omega$ in a certain region of the parameters, but its applicability conditions are violated in the limit $\omega \longrightarrow 0$, which is of main interest.

(ii) In fact, Sadovskiĭ [5] and Cardy [10] obtained the singularity[2]

$$\langle G_{E_1} G_{E_2} \rangle \sim \frac{1}{E_1 - E_2}, \quad (6)$$

For $E_1 = E + \omega + i\delta$ and $E_2 = E - i\delta$, this singularity corresponds to Eq. (3). However, such a singularity must be absent when $E_1$ and $E_2$ have the imaginary additions of the same sign (this case corresponds to the averages $\langle G^R G^R \rangle$ and $\langle G^A G^A \rangle$). It is unclear from [5, 10], which mechanism ensures the disappearance of the singularity in the last case.

(iii) Singularity (6) is in agreement with singularity (3) if $E_1$ and $E_2$ are treated as the bare energies. However, according to the conventional paradigm (see, e.g., the calculation of conductivity in [11]), the renormalized energies that contain terms with finite damping $\pm i\Gamma$ rather than infinitely small imaginary additions $\pm i\delta$ should be used in such cases. The correct procedure of instanton calculations [12] (for more details, see [1, 13]) also requires the transition to the renormalized energy, because otherwise the instanton contribution diverges. The replacement of $i\delta$ by $i\Gamma$ would lead to the disappearance of the diffusion pole in the localized phase[3], which would result in large difficulties for the theory, because the initial Anderson localization criterion [16, 17] is called in question (see discussion in [5, 6, 10]).

(iv) The momentum dependence of the results reported in [5, 10] is in disagreement with the dependence following from the direct analysis of the Bethe–Salpeter equation [3].

(v) The calculations performed in [5, 10] are deterministic and give no indications to the origin of the self-consistent treatment for the diffusion coefficient; the necessity of such a procedure seems to be reasonable from the physical concepts (see above).

It will be shown in Section 5 that problems (i) and (ii) are of technical character and can be solved by modifying the calculation procedure. In particular, it is mathematically more correct to apply the instanton method to calculate high orders of perturbation theory and to obtain a singular contribution as a result of their summation. This procedure provides a clear interpretation of the arising expressions and removes the problem of imaginary additions. The problem of the inapplicability of the saddle-point approximation can be overcome by the correct integration over the soft mode.

Problems (iii)–(v) are deeper: they are not removed even if the problem is extremely simplified by taking zero-dimensional limit (which corresponds to the strong disorder limit in arbitrary dimension). On the contrary, they are extremely sharpened in this case and become paradoxes. The main paradox is specified by the following statements: (i) the $1/\omega$ singularity is determined by high orders of perturbation theory, (ii) the high-order behaviors for $\Phi^{RA}$ and $U^{RA}$ are the same, and (iii) the $1/\omega$ singularity exists in the quantity $\Phi^{RA}$, whereas it is absent in the quantity $U^{RA}$.[4] Analysis of this paradox is the main aim of this work. Its solution leads to the following conclusions. The instanton approach determines only the general structure of the diffusion pole and makes it possible to obtain the $1/(\omega + 2i\gamma)$ singularity, whereas the parameter $\gamma$ remains indefinite and should be determined with the inclusion of information from low orders of perturbation theory. It is individual for a quantity under consideration and is generally a function of the momenta. It is physically expected that $\gamma$ for the quantity $\Phi^{RA}$ in the localized phase is on the order of $\omega$. Owing to this fact, the $1/\omega$ singular contribution is recovered, but its momentum dependence changes. The parameter $\gamma$ in the metallic phase is finite in correspondence with the finite diffusion coefficient and must be determined by a certain self-consistent procedure. This behavior confirms the Vollhardt–Wölfle scenario at the fundamental level.

The cross verification of all statements is of current interest due to the existence of the paradox. Moreover, the investigation of the zero-dimensional case is useful for the combinational analysis of the diagrams [18, 19], construction of expansions in the dimension of space [20], and use in interpolation schemes [21]. For this

---

[1] The inclusion of the spatial dispersion of $D(\omega, q)$ is of general physical interest and of significant importance for the satisfaction of the Ward identity [2],

$$\Delta\Sigma_k(\mathbf{q}) = \frac{1}{N}\sum_{k_1} U_{kk_1}^{RA}(\mathbf{q})\Delta G_{k_1}(\mathbf{q})$$

where $\Delta G_k(\mathbf{q}) \equiv G_{k+q/2}^R - G_{k-q/2}^A$, $\Delta\Sigma_k(\mathbf{q}) \equiv \Sigma_{k+q/2}^R - \Sigma_{k-q/2}^A$, and $\Sigma_k$ is the self energy, because the summation with respect to $\mathbf{k}_1$ involving $D(\omega, \mathbf{k} + \mathbf{k}_1)$ must ensure the cancellation of the $1/\omega$ singularity on the right-hand side.

[2] The notation $G_E$ is used for the Green's function when the sign of the imaginary addition to the energy is of no importance.

[3] Such statement was made in recent works [14], but it is incorrect in our opinion [15].

[4] The last statement is specific for the zero-dimensional case and is associated with the absence of the momentum integration in the Ward identity (see Footnote 1).

reason, the approaches based on the functional integration and exact diagrammatic expansions are given in Sections 2 and 3, respectively. The physical interpretation is presented in Section 4. The calculation procedure for high orders of perturbation theory is considered in Section 5 and their summation is discussed in Section 6. The solution to the formulated paradox is given in Section 7. Finally, the modification of instanton calculations performed in [5, 10] is presented in Section 8 on the basis of the performed analysis.

## 2. ZERO-DIMENSIONAL LIMIT IN THE FUNCTIONAL INTEGRAL

The average Green's function of the disordered system and the quantity $\Phi^{RA}$ can be represented in the form of the functional integrals [22]

$$\langle G(x, x') \rangle = -\int D\varphi \, \varphi_\alpha(x) \varphi_\alpha(x')$$
$$\times \exp\left\{-\int d^d x \left[\frac{1}{2}(\nabla\varphi)^2 + \frac{1}{2}\kappa^2\varphi^2 + \frac{1}{4}g\varphi^4\right]\right\}, \quad (7)$$

where $\kappa^2 = -E \pm i\delta$ and

$$\Phi^{RA}(x_1, x_2, x_3, x_4) = \int D\varphi D\phi \, \varphi_\alpha(x_1)\varphi_\alpha(x_2) \qquad (8)$$
$$\times \phi_\beta(x_3)\phi_\beta(x_4) \exp(-S\{\varphi, \phi\}),$$

Here,

$$S\{\varphi, \phi\} = \int d^d x \left\{\frac{1}{2}(\nabla\varphi)^2 + \frac{1}{2}(\nabla\phi)^2 + \frac{1}{2}\kappa_1^2\varphi^2 \right.$$
$$\left. + \frac{1}{2}\kappa_2^2\phi^2 + \frac{1}{4}g(\varphi^2 + \phi^2)^2\right\},$$

where

$$g = -W^2/2, \quad \kappa_1^2 = -(E + \omega + i\delta) \equiv \kappa^2 - i\Delta,$$
$$\kappa_2^2 = -(E - i\delta) \equiv \kappa^2 + i\Delta. \quad (9)$$

$\varphi$ and $\phi$ are the $m$- and $n$-component vectors, respectively, and the passage to the limit $m, n \longrightarrow 0$ is performed at the end of calculations.[5] Formulas (7) and (8) correspond to the usual Anderson model with the Gaussian distribution of the energies of the sites

$$P(V) = \frac{1}{\sqrt{2\pi W^2}} \exp\left\{-\frac{V^2}{2W^2}\right\}, \quad (10)$$

which is considered near the lower edge of the band and, in the continual limit, is equivalent to the Schrödinger equation with the $\delta$ correlated random potential.

Let us treat the functional integral as a multiple integral on a lattice. For the passage to the zero-dimensional limit, let us consider a system spatially limited in all the directions at a sufficiently small scale. For this system, the coordinate dependences $\varphi(x)$ and $\phi(x)$ can be disregarded and the gradient terms can be omitted in Eqs. (7) and (8). Taking a sufficiently sparse lattice, one can assume that only one its site is located inside the system. In this case,

$$\langle G \rangle = -\int_0^\infty \varphi d\varphi \exp\left\{-\frac{1}{2}\kappa^2\varphi^2 - \frac{1}{4}g\varphi^4\right\}, \quad (11)$$

where it is taken into account that

$$\int d^n\varphi \, \varphi_\alpha\varphi_\alpha = S_n\int_0^\infty \varphi^{n-1} d\varphi \frac{\varphi^2}{n} \longrightarrow \int_0^\infty \varphi d\varphi \quad (12)$$

in the limit $n \longrightarrow 0$, where $S_n$ is the area of the unit sphere in the $n$-dimensional space, and, similarly,

$$\Phi^{RA} = \int_0^\infty \varphi d\varphi \int_0^\infty \phi d\phi \qquad (13)$$
$$\times \exp\left\{-\frac{1}{2}\kappa_1^2\varphi^2 - \frac{1}{2}\kappa_2^2\phi^2 - \frac{1}{4}g(\varphi^2 + \phi^2)^2\right\}.$$

The power expansion in $g$ provides the series

$$\langle G \rangle = -\frac{1}{\sqrt{\pi}\kappa^2} \sum_{N=0}^\infty \left(-\frac{4}{\kappa^4}\right)^N \Gamma\left(N + \frac{1}{2}\right) g^N, \quad (14)$$

$$\Phi^{RA} = \frac{1}{\sqrt{\pi}(\kappa_1^2 - \kappa_2^2)} \sum_{N=0}^\infty (-4)^N \Gamma\left(N + \frac{1}{2}\right) \qquad (15)$$
$$\times [(\kappa_2^2)^{-(2N+1)} - (\kappa_1^2)^{-(2N+1)}] g^N,$$

where the binomial formula and the multiplication theorem for the gamma function are used [23].

## 3. DIAGRAMMATIC EXPANSIONS IN THE ZERO-DIMENSIONAL CASE

The Green's functions in the zero-dimensional case are independent of the coordinates or correspond to zero momentum in the momentum representation. The diagrammatic expansions (see Fig. 1) have the usual form, but do not contain coordinate or momentum integrals. All the diagrams of the same order at $\kappa_1^2 = \kappa_2^2$ are equal to each other and the perturbation series can be constructed on the basis of the combinational analysis of the diagrams. At the same time, the combinational analysis of the diagrams can be performed by means of

---
[5] The summation over repeated subscripts $\alpha$ and $\beta$ is not implied in Eqs. (7) and (8).

"functional integrals" (11) and (13) [18, 19]. The actual diagrams are shown in Fig. 1, where the dashed line corresponds to the factor $W^2$, the thin solid line means the bare Green's function

$$G^0(E) = \frac{1}{E} = -\frac{1}{\kappa^2}, \qquad (16)$$

and the thick solid line stands for the total Green's function

$$G(E) \equiv \langle G_E \rangle = \frac{1}{\tilde{E}} = \frac{1}{E - \Sigma} = -\frac{1}{\tilde{\kappa}^2}, \qquad (17)$$

where $\Sigma \equiv \Sigma_R = -\Delta E + i\Gamma$ for $G^R$ (and similarly for $G^A$). In particular, the series in Fig. 1a has the form

$$G(E) = \frac{1}{E} \sum_{N=0}^{\infty} G_N \left(\frac{W^2}{E^2}\right)^N,$$

$$G_N = (2N-1)!! = 2^N \frac{\Gamma(N+1/2)}{\sqrt{\pi}}, \qquad (18)$$

because the number of diagrams of the $N$th order is equal to $(2N-1)!!$ [the first of $2N$ vertices is connected to one of the $2N-1$ remaining vertices, the first free vertex is connected to one of the $2N-3$ remaining vertices, etc.). The self-energy $\Sigma$ (see Fig. 1b) can be expanded in both bare (see Fig. 1c) and renormalized (see Fig. 1d) Green's functions

$$\Sigma(E) = E \sum_{N=1}^{\infty} L_N \left(\frac{W^2}{E^2}\right)^N$$

$$= \tilde{E} \sum_{N=1}^{\infty} K_N \left(\frac{W^2}{\tilde{E}^2}\right)^N, \qquad (19)$$

where the sequences $L_N$ and $K_N$ are determined by the recurrence relations

$$L_N = \sum_{K=0}^{N-1} G_{N-K} L_K, \quad L_0 = 1,$$

$$K_{N+1} = N \sum_{M=0}^{N-1} K_{M+1} K_{N-M}, \quad K_0 = 1. \qquad (20)$$

The first relation follows from the equality $G(E - \Sigma) = 1$ as a result of the multiplication of series and the second relation was obtained in [19]. It is interesting that the sequence $K_N$ also determines the expansions for the quantities $\Phi$ and $U$ (which are analogues of $\Phi^{RA}$ and $U^{RA}$ at $\kappa_1^2 = \kappa_2^2$) in the renormalized Green's function [19]

$$\Phi = \frac{1}{\tilde{E}^2} \sum_{N=0}^{\infty} K_{N+1} \left(\frac{W^2}{\tilde{E}^2}\right)^N, \qquad (21)$$

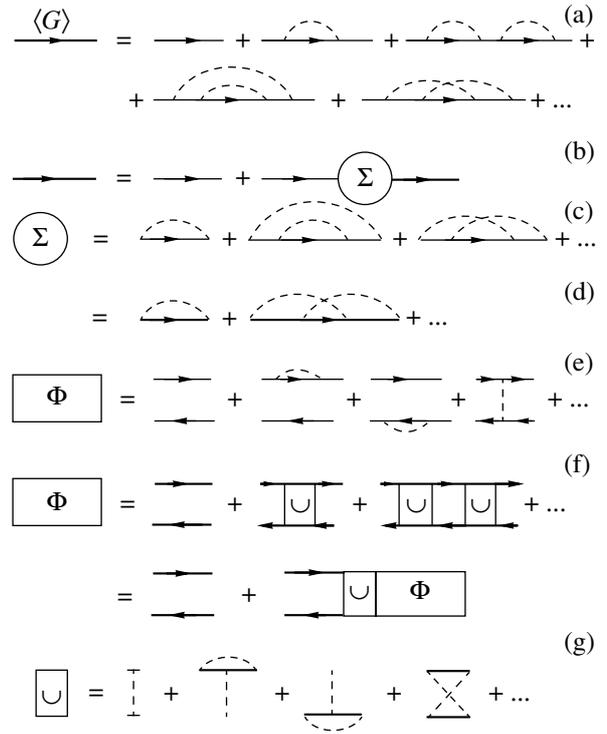

**Fig. 1.** Diagrammatic expansions for the electron in a random potential.

$$U = \tilde{E}^2 \sum_{N=1}^{\infty} (2N-1) K_N \left(\frac{W^2}{\tilde{E}^2}\right)^N. \qquad (22)$$

The first relation follows from the Ward identity relating $\Sigma$ and $\Phi$ and the second relation follows from the fact that all the diagrams for $U$ (see Fig. 1g) can be obtained from the diagrams for $\Sigma$ (see Fig. 1d) by successive cutting along one of the $2N-1$ internal $G$ lines. Similarly, the expansion of $\Phi$ in the bare Green's functions,

$$\Phi = \frac{1}{E^2} \sum_{N=0}^{\infty} (2N+1)!! \left(\frac{W^2}{E^2}\right)^N, \qquad (23)$$

follows from the fact that the diagrams for $\Phi$ (see Fig. 1e) are obtained from the diagrams for $G$ (see Fig. 1a) by cutting along one of the $2N + 1$ electron lines. From the method of constructing expansions (23) and (24), it is clear how they are generalized to the case $\kappa_1^2 \neq \kappa_2^2$:

$$\Phi^{RA} = \sum_{N=0}^{\infty} (2N-1)!! W^{2N}$$

$$\times \sum_{K=0}^{2N+1} (G_0^R)^K (G_0^A)^{2N+2-K}, \qquad (24)$$

$$U^{RA} = \sum_{N=1}^{\infty} K_N W^{2N} \qquad (25)$$

$$\times \sum_{K=1}^{2N-1} (G^R)^{K-1} (G^A)^{2N-1-K}.$$

It is easy to verify that expressions (18) and (24) are equivalent to Eqs. (14) and (15) presented in the preceding section.

## 4. PHYSICAL INTERPRETATION AND EXACT RESULTS

The zero-dimensional case physically corresponds to the Anderson model at one site of the lattice. The exact Green's function

$$G^R(E) = \frac{1}{E - V + i0} \qquad (26)$$

after the trivial averaging has the form

$$\langle G^R(E) \rangle = \int_{-\infty}^{\infty} \frac{P(V) dV}{E - V + i0} \equiv \frac{1}{E + \Delta E + i\Gamma}. \qquad (27)$$

The density of states is obviously equal to

$$\nu(E) = -\frac{1}{\pi} \mathrm{Im} \langle G^R(E) \rangle = P(E), \qquad (28)$$

and the renormalized energy and damping are given by the expressions

$$E + \Delta E = \frac{f(E)}{[\pi \nu(E)]^2 + [f(E)]^2},$$
$$\Gamma = \frac{\pi \nu(E)}{[\pi \nu(E)]^2 + [f(E)]^2}, \qquad (29)$$

where

$$f(E) = \mathrm{P.v.} \int_{-\infty}^{\infty} \frac{P(V) dV}{E - V}. \qquad (30)$$

The quantity $\Phi^{RA}$ is specified by the expression

$$\Phi^{RA} = \langle G^R(E + \omega) G^A(E) \rangle$$
$$= \int_{-\infty}^{\infty} \frac{1}{E - V + \omega + i0} \frac{1}{E - V - i0} P(V) dV \qquad (31)$$

and, at $\omega \longrightarrow 0$, has the singularity

$$\Phi^{RA} = \frac{2\pi P(E)}{-i\omega}, \qquad (32)$$

obviously coinciding with Eq. (3). From the Bethe–Salpeter equation (see Fig. 1f)

$$\Phi^{RA} = G^R G^A [1 + U^{RA} \Phi^{RA}], \qquad (33)$$

the vertex $U^{RA}$ is obtained in the form

$$U^{RA} = \frac{1}{G^R G^A} + O(\omega). \qquad (34)$$

The same result can be obtained from the Ward identity (see Footnote 1)

$$\Delta \Sigma = U^{RA} \Delta G. \qquad (35)$$

For Gaussian distribution (10), Eqs. (27) and (31) are equivalent to Eqs. (11) and (13), respectively. Indeed, replacement of the gamma function in Eqs. (14) and (15) to the determining integral and the summation of the appearing geometric progression (which corresponds to the Borel summation in the theory of divergent series [24]) provide the expressions

$$\langle G \rangle = \frac{1}{\kappa^2 \sqrt{\pi}} \int_0^{\infty} dx \, x^{-1/2} e^{-x} \left[1 - \frac{4|g|x}{\kappa^4}\right]^{-1}, \qquad (36)$$

$$\Phi^{RA} = \frac{1}{\sqrt{\pi}(\kappa_1^2 - \kappa_2^2)} \int_0^{\infty} dx \, x^{-1/2}$$
$$\times e^{-x} \left[\frac{\kappa_2^2}{\kappa_2^4 - 4|g|x} - \frac{\kappa_1^2}{\kappa_1^4 - 4|g|x}\right]^{-1}, \qquad (37)$$

which, after the substitution $x = V^2/4|g|$, are reduced to Eqs. (27) and (31), respectively.

## 5. HIGH ORDERS OF PERTURBATION THEORY

The possibility of constructing exact diagrammatic expansions is specific for the zero-dimensional case. The role of high orders, which can be analyzed in arbitrary dimension, is more interesting. High-order asymptotic expressions for expansions presented in Section 3 are determined by the two results

$$L_N \approx (2N-1)!!, \quad K_N \approx \frac{1}{e}(2N-1)!!, \qquad (38)$$

The first of them follows from the fact that expansions (18) and (19) in bare energies are related to each other due to the relation $G(E) = 1/(E - \Sigma)$. With the use of the algebra of the factorial series [1, 13] for a power of –1,

$$(A_0 + A_1 g + \ldots + A_N g^N + \ldots)^{-1}$$
$$= A_0^{-1} - A_0^{-2} A_1 g - \ldots - A_0^{-2} A_N g^N - \ldots, \qquad (39)$$

it is easy to show that $L_N \approx G_N$ in high orders. The result for $K_N$ was obtained by Kuchinskiĭ and Sadovskiĭ [19]: the expression $K_N = \mathrm{const}(2N-1)!!$ follows from recursion relations (21) and the equality $\mathrm{const} = 1/e$ was found numerically.

A general method for investigating high orders is the Lipatov method [25] based on the fact that the coefficients $F_N$ of the expansion of the function $F(g)$ in the power series in $g$ are expressed in terms of $F(g)$ in the form of the contour integral. In particular, for Eq. (11),

$$G_N = -\oint_C \frac{dg}{2\pi g}\int_0^\infty \varphi d\varphi \times \exp\left\{-\frac{1}{2}\kappa^2\varphi^2 - \frac{1}{4}g\varphi^4 - N\ln g\right\} \quad (40)$$

where $C$ is the contour around the point $g = 0$ and the integral at large $N$ values is determined by the contribution from the saddle-point configuration

$$\varphi_c^2 = \frac{4N}{\kappa^2}, \quad g_c = -\frac{\kappa^4}{4N}, \quad (41)$$

which gives the result

$$G_N = -\frac{1}{\sqrt{\pi\kappa^2}}\left(-\frac{4}{\kappa^4}\right)^N \Gamma\left(N + \frac{1}{2}\right), \quad N \gg 1, \quad (42)$$

which is exact in this case [see Eq. (14)].

Similarly for Eq. (13),

$$[\Phi^{RA}]_N = \oint_C \frac{dg}{2\pi g}\int_0^\infty \varphi d\varphi \int_0^\infty \phi d\phi$$
$$\times \exp\left\{-\frac{1}{2}\kappa^2(\varphi^2 + \phi^2)\right. \quad (43)$$
$$\left. -\frac{1}{2}i\Delta(\varphi^2 - \phi^2) - \frac{1}{4}g(\varphi^2 + \phi^2)^2 - N\ln g\right\}.$$

The saddle points at $\Delta = 0$ form a continuous set and lie on the circle $\varphi^2 + \phi^2 = 4N/\kappa^2$. The saddle points at $\Delta \ne 0$ are discrete points ($\varphi^2 = 0$, $\phi^2 = 4N/\kappa^2$) and ($\varphi^2 = 4N/\kappa^2$, $\phi^2 = 0$) (see Fig. 2). The calculation of the contribution from the latter points corresponds to the approach used in [5, 10]. In this case, the saddle-point approximation is inapplicable in the limit $\Delta \longrightarrow 0$, because change in the action along the circle $\varphi^2 + \phi^2 = 4N/\kappa^2$ for small $\Delta$ values is very small and this is a typical soft mode: integration along this circle should be performed exactly rather than in the saddle-point approximation.

The general idea for treating soft modes [24, 26] is that the instanton is sought as the extremum of the action under an additional condition (constraint), which fixes the degree of freedom corresponding to the soft mode; after that, the integration over this degree of freedom is performed. In this case, it is convenient to take the constraint in the form

$$\varphi^2 - \phi^2 = \xi = \text{const}, \quad (44)$$

so that the extremum of the following function should be sought:

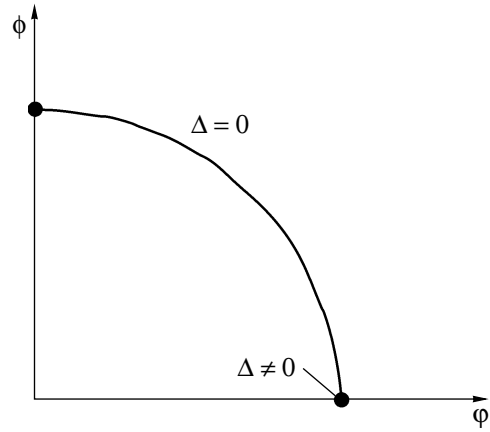

**Fig. 2.** Saddle points for integral (44).

$$\frac{1}{2}\kappa^2(\varphi^2 + \phi^2) + \frac{1}{4}g(\varphi^2 + \phi^2)^2$$
$$-\frac{1}{2}i\Delta(\varphi^2 - \phi^2) - \mu(\varphi^2 - \phi^2 - \xi), \quad (45)$$

where $\mu$ is the Lagrange multiplier. The extremum exists at $\mu = -i\Delta/2$, when function (45) corresponds to the action of the symmetric problem ($\Delta = 0$).

The formal procedure is the introduction of the decomposition of unity,

$$1 = \int_{-\infty}^\infty d\xi \delta(\xi - \varphi^2 + \phi^2), \quad (46)$$

that transforms integral (13) to the form

$$\Phi^{RA} = \int_0^\infty \varphi d\varphi \int_0^\infty \phi d\phi \int_{-\infty}^\infty d\xi \delta(\xi - \varphi^2 + \phi^2)$$
$$\times \exp\left\{-\frac{1}{2}\kappa^2(\varphi^2 + \phi^2) - \frac{1}{4}g(\varphi^2 + \phi^2)^2 + \frac{1}{2}i\Delta\xi\right\}. \quad (47)$$

In terms of the polar coordinates $r$ and $\theta$ introduced through the relation $\varphi = r\cos\theta$ and $\phi = r\sin\theta$,

$$\Phi^{RA} = \int_0^{\pi/2} d\theta \sin 2\theta \int_{-\infty}^\infty d\xi e^{\frac{1}{2}i\Delta\xi} \int_0^\infty r^3 dr$$
$$\times \exp\left\{-\frac{1}{2}\kappa^2 r^2 - \frac{1}{4}gr^4\right\}\delta(\xi - r^2\cos 2\theta), \quad (48)$$

where the integral over $r$ (disregarding the $\delta$ function term) is similar to Eq. (11) and its high-order expansion coefficients are determined by the saddle point $r_c = \sqrt{4N/\kappa^2}$. For small $\Delta$ values, deviations of $r$ from $r_c$ in the Dirac $\delta$ function can be neglected, because they are insignificant as compared to change in θ. This allows one to factorize the dependence on $\Delta$:

$$\Phi^{RA}(\Delta) = \Phi^{RA}(0) \int_0^{\pi/2} 2d\theta \sin 2\theta \qquad (49)$$
$$\times \exp\left\{\frac{1}{2}i\Delta r_c^2 \cos 2\theta\right\}.$$

After the calculation of the expansion coefficients for $\Phi^{RA}(0)$, the resulting expression has the form

$$[\Phi^{RA}(\Delta)]_N = \frac{2}{\sqrt{\pi}\kappa^4}\left(-\frac{4}{\kappa^4}\right)^N$$
$$\times \Gamma\left(N + \frac{3}{2}\right)\frac{e^{i\Delta\lambda N} - e^{-i\Delta\lambda N}}{2i\Delta\lambda N}, \quad \lambda = \frac{2}{\kappa^2}. \qquad (50)$$

For small $\Delta$ values, this expression coincides with the asymptotic form of the coefficients in Eq. (15). The described method can be generalized on arbitrary dimension of space (see Section 8).

## 6. WHAT CAN BE EXTRACTED FROM HIGH ORDERS?

The contribution from high orders of perturbation theory is often substantial. This is associated with the relation [1, 13]

$$\text{Im} \sum_{N = N_0}^{\infty} ca^N \Gamma(N + b) g^N e^{i\delta N}$$
$$= \frac{\pi c}{(ag)^b} e^{-1/ag}, \quad ag > 0, \qquad (51)$$

which is obtained by Borel summation of the series on the left-hand side. In the limit $\delta \longrightarrow 0$, the result is independent of $N_0$, because the imaginary part of each term of the series (and the sum of the finite number of these terms) is infinitely small. Result (51) is valid for finite $\delta$ values under the condition $N_0 \ll 1/\delta$, which is usually satisfied, because $\delta$ is usually exponentially small. The form $ca^N\Gamma(N + b)$ is typical for the high-order asymptotic expressions obtained by the Lipatov method [24]. Corrections to it have the form of a regular expansion in $1/N$, which provides a regular expansion in $g$ after summation according to Eq. (51). For this reason, result (51) is constructive at $g \ll 1$. In this case, the real part of the series is well approximated by several first terms. The applicability of such results for the expansions considered in Section 3 is restricted by the region $|E| \gg W$. The nonperturbative results such as $\exp(-\text{const}/g)$ are usually obtained by the instanton method, but relation (51) makes it possible to reproduce them directly from perturbation theory.

The summation of high orders for $\langle G \rangle$ [see Eq. (18)] allows one to obtain the fluctuation tail of the density of states

$$\nu(E) = -\frac{1}{\pi}\text{Im}\langle G^R(E)\rangle = -\frac{1}{\pi}\text{Im}\frac{1}{E + i\delta}$$
$$\times \sum_{N = N_0}^{\infty} 2^N\frac{\Gamma(N + 1/2)}{\sqrt{\pi}}\left(\frac{W}{E + i\delta}\right)^{2N} = \frac{e^{-E^2/2W^2}}{\sqrt{2\pi W^2}} \qquad (52)$$

according to Eq. (28) for distribution (10). The summation of series (19) for $\Sigma$ yields the following expression for damping $\Gamma$:

$$\Gamma = -\text{Im}\Sigma^R = -\text{Im}(E + i\delta)$$
$$\times \sum_{N = N_0}^{\infty} 2^N\frac{\Gamma(N + 1/2)}{\sqrt{\pi}}\left(\frac{W}{W + i\delta}\right)^{2N} = \frac{\pi E^2 e^{-E^2/2W^2}}{\sqrt{2\pi W^2}}, \qquad (53)$$

which can also be obtained from Eq. (29) at $|E| \gg W$, when $f(E) \approx 1/E \gg \pi\nu(E)$.

The summation of the series with coefficients (50) provides the $1/\omega$ singularity:

$$[\Phi^{RA}]_{\text{sing}} = \sum_{N = N_0}^{\infty} [\Phi^{RA}]_N g^N = \frac{2\pi}{-i\omega}\frac{e^{-E^2/2W^2}}{\sqrt{2\pi W^2}} \qquad (54)$$

in agreement with Eq. (32). It is very important that the exponentials in the numerator in Eq. (50) have imaginary arguments with opposite signs. For this reason, the real parts of the series cancel, whereas the imaginary parts are summed and the $1/\omega$ singularity appears in (50). If the imaginary additions to the energy in Eq. (9) had the same signs (which corresponds to averages such as $\langle G^R G^R \rangle$ or $\langle G^A G^A \rangle$), the arguments of the exponentials in the numerator in Eq. (50) would have the form $i\Delta_1\lambda N$ and $i\Delta_2\lambda N$ with $\text{Re}\Delta_1 \cdot \text{Re}\Delta_2 > 0$. This circumstance would lead to the cancellation (in the leading order in $\Delta$) of both real and imaginary parts of two series of form (51) and to the removal of the singularity.

The contribution from high orders to sum (25) for $U^{RA}$ is calculated similarly: with $\kappa_1^2 = -\bar{E}e^{-i\beta}$ and $\kappa_2^2 = -\bar{E}e^{i\beta}$,

$$U^{RA}G^R G^A = \sum_{N = N_0}^{\infty} 2^N\frac{\Gamma(N + 1/2)}{\sqrt{\pi}e}\left(\frac{W}{\bar{E}}\right)^{2N}$$
$$\times \frac{e^{i\beta(2N-1)} - e^{i\beta(2N-1)}}{e^{i\beta} - e^{-i\beta}} \approx \frac{\pi\bar{E}P(\bar{E})}{e\beta}, \qquad (55)$$

The substitution of $\beta = \Gamma/\bar{E}$ and Eq. (53) provides the result $U^{RA}G^R G^A \approx 1$ coinciding with Eq.(34).

It is easily seen that all the physical results in the region of the fluctuation tail ($|E| \gg W$) (see Section 4) are determined by high orders. The contribution from these orders in all cases prevails over the contribution from the first terms of the series, which are treated in the asymptotic sense [24].

## 7. PARADOX AND ITS SOLUTION

Unfortunately, the results of the summation for $\Phi^{RA}$ and $U^{RA}$ [see Eqs. (54) and (55)] coincide with the exact results (see Section 4) only incidently. According to Bethe–Salpeter equation (34),

$$U^{RA} = \frac{1}{G^R G^A} - \frac{1}{\Phi^{RA}}, \quad (56)$$

i.e., the series for $\Phi^{RA}$ and $U^{RA}$ are determined by the mutually inverse functions and their high-order expansion coefficients almost coincide with each and differ only in a constant factor [see Eq. (39)]. From Eqs. (24,25) it is easy to obtain the expansion of $U^{RA}$ in the bare Green's functions and the expansion of $\Phi^{RA}$ in the renormalized Green's functions. The summation of these expansions provides the results

$$\Phi^{RA} = \frac{\pi P(\bar{E})}{e\beta\bar{E}}, \quad U^{RA} G^R G^A = \frac{2\pi E^2 P(E)}{-i\omega}, \quad (57)$$

which are in sharp disagreement with the results given in Eqs. (54), (55), and Section 4.

The main of arising paradoxes was formulated in Section 1. The $1/\omega$ singularity is determined by high orders (see Section 6). The high-order behaviors for $\Phi^{RA}$ and $U^{RA}$ are the same according to Eqs. (56) and (39). However, $\Phi^{RA}$ has the singularity, whereas $U^{RA}$ does not have it according to Eqs. (32) and (34), respectively.

It is convenient to begin to analyze the paradox with the derivation of the second of results (38), where the factor $1/e$ was determined in [19] only numerically. As will be seen, this provides a key to the problem. As seen from Eq. (19), $K_N$ is the number of the "skeleton" diagrams of the $N$th order for the self-energy $\Sigma$ (see Fig. 1d), whereas $L_N$ is the number of the diagrams for $\Sigma$, where the self-energy insertions are not excluded (see Fig. 1c). As mentioned above, for high orders, $L_N \approx (2N-1)!!$ due to Eq. (39) and $K_N = \text{const}(2N-1)!!$ due to Eq. (20), where const should be determined. It is easy to understand that $L_N$ includes

(i) the $K_N$ skeleton diagrams;

(ii) the diagrams that are obtained from the skeleton diagrams of the (N – 1)th order; the number of such diagrams is $K_{N-1} \approx K_N/2N$, but approximately $2N$ one-loop insertions are possible in the internal lines; therefore, the total number of these diagrams is also equal to $K_N$;

(iii) the diagrams that are obtained from the skeleton diagrams of the (N – 2)th order; the number of such diagrams is $K_{N-2} \approx K_N/(2N)^2$, but approximately $(2N)^2/2!$ one-loop insertions are possible and this provides $K_N/2!$ diagrams; in addition, approximately $2N$ two-loop insertions are possible and this provides about $K_N/N$ diagrams, but this number is insignificant for large $N$ values;

(iv) the diagrams that are obtained from the skeleton diagrams of the (N – 3)th order; the number of such diagrams is $K_{N-3} \approx K_N/(2N)^3$, but approximately $(2N)^3/3!$ one-loop insertions are possible and this provides $K_N/3!$ diagrams; in addition, approximately $2N$ three-loop insertions or about $(2N)^2$ combined (two- + one-loop) insertions are possible, but this contribution is insignificant for large $N$ values; etc.

It is seen that the difference between $L_N$ and $K_N$ in the leading order is determined by the one-loop insertions; therefore,

$$L_N = K_N(1 + 1/1! + 1/2! + 1/3! + \ldots) = eK_N, \quad (58)$$

which was to be proved.

The property revealed above is a particular case of a more general statement [12] that the leading Lipatov asymptotic form is sensitive to the renormalization of the energy (or "mass" in the field-theory terms) only at the one-loop level and only in the form of a constant factor. More precisely, let the Lipatov asymptotic form in terms of the bare energy $E$ is expressed as

$$c(E)a(E)^N \Gamma(N+b) \quad (59)$$

where the parameter $b$ is determined only by the number of the external lines and zeroth modes and is energy independent [24]. Then, in terms of the renormalized energy $\tilde{E}$, it has the form

$$k(\tilde{E})c(\tilde{E})a(\tilde{E})^N \Gamma(N+b), \quad (60)$$

where $k$ is independent of $N$ and $c(\tilde{E})$ and $a(\tilde{E})$ are the same functions as in Eq. (59).

Indeed, the transition from the bare energy to the renormalized energy, $\kappa^2 = \tilde{\kappa}^2 + \Sigma(0, \tilde{\kappa}^2)$, in functional integrals (7) and (8) reduces to the change of variables

$$\kappa^2 = \tilde{\kappa}^2 + a_1(\tilde{\kappa}^2)g + a_2(\tilde{\kappa}^2)g^2 + a_3(\tilde{\kappa}^2)g^3 + \ldots \quad (61)$$

In the calculation of high orders, the saddle-point configuration in which $\varphi_c^2 \sim \phi_c^2 \sim N$ and $g_c \sim 1/N$ [cf. Eq. (41)] is substantial in the functional integral. For this reason, in Eq. (7),

$$\exp(-S\{\varphi_c, \kappa^2\}) = \exp(-S\{\varphi_c, \tilde{\kappa}^2\} - A),$$
where
$$A = \frac{1}{2}a_1(\tilde{\kappa}^2)g_c \int d^d x \varphi_c^2(x) + O(1/N); \quad (62)$$

i.e., the transition from $\kappa^2$ to $\tilde{\kappa}^2$ conserves the functional form of the saddle-point contribution and leads only to the appearance of the factor $e^{-A}$ that is independent of $N$ and is determined by the coefficient $a_1(\tilde{\kappa}^2)$, i.e., by the one-loop contribution.[6] Hence, the Lipatov asymptotic form does not change if the coefficients $a_2(\tilde{\kappa}^2)$, $a_3(\tilde{\kappa}^2)$, etc. are set to zero. At the same time, the complete renormalization of energy qualitatively differs from the one-loop renormalization. In the first case, the damping $\Gamma$ in $\tilde{\kappa}^2 = -E - \Delta E - i\Gamma$ is the same as in the average Green's function and is finite for all $E$ values [see Eq. (29)].[7] In the second case, the imaginary part of $\tilde{\kappa}^2$ is infinitely small in the entire region of the fluctuation tail. Moreover, an arbitrary change in variables specified by Eq. (61) with arbitrarily chosen coefficients $a_2(\tilde{\kappa}^2)$, $a_3(\tilde{\kappa}^2)$, etc. makes it possible to arbitrarily change the imaginary part of $\tilde{\kappa}^2$ without change in the functional form of the Lipatov asymptotic form.[8]

According to the above consideration, the instanton method allows one to find only the general structure of the singularities for $\Phi^{RA}$ and $U^{RA}$:

$$\Phi^{RA} \sim \frac{1}{\omega + 2i\gamma}, \quad U^{RA} \sim \frac{1}{\omega + 2i\tilde{\gamma}}, \quad (63)$$

where the parameters $\gamma$ and $\tilde{\gamma}$ remain indefinite and cannot be determined in the framework of the instanton method, neither in the leading saddle-point approximation nor in the case of the inclusion of the finite-order corrections to it. They must be determined with the inclusion of information from low orders of perturbation theory. The parameters $\gamma$ and $\tilde{\gamma}$ in the zero-dimensional case are unambiguously determined with the use of Ward identity (35) and Bethe–Salpeter equation (33) by the expressions

$$\gamma = 0, \quad \tilde{\gamma} = \Gamma, \quad (64)$$

which obviously demonstrate that the general principles responsible for the preferable use of the bare energy or renormalized energy are absent. Thus, the parameters $\gamma$ and $\tilde{\gamma}$ are different and generally depend on the momenta. The problem of determining them for arbitrary dimension of space is open. Since the singularities of $\Phi^{RA}$ and $U^{RA}$ are determined by the same diffusion coefficient [see Eqs. (2) and (4)], it is confirmed that this coefficient should be calculated with the self-consistent treatment used in the Vollhardt–Wölfle theories [1, 2].

---

[6] The values $a_2(\tilde{\kappa}^2)$, $a_3(\tilde{\kappa}^2)$, etc. become significant when the corrections of the orders $1/N$, $1/N^2$, etc. are successively taken into account.

[7] This is also the case for arbitrary dimension of space [1, 13, 27].

[8] Note that an arbitrary one-loop renormalization of the energy [i.e., the correct $a_1(\tilde{\kappa}^2)$ value] at $d \geq 2$ is fundamentally necessary, because it eliminates the ultraviolet divergence in the parameter $c$ of the Lipatov asymptotic form [1, 12, 13].

## 8. INSTANTON CALCULATIONS IN THE GENERAL CASE

Let us show how the instanton calculations performed in [5, 10] should be modified in view of the above analysis. The introduction of the decomposition of unity

$$1 = \int_{-\infty}^{\infty} d\xi \delta(\xi - \int d^d x \boldsymbol{\varphi}^2(x) + \int d^d x \boldsymbol{\phi}^2(x)), \quad (65)$$

which is similar to Eq. (46), into integral (8) reduces it to the form

$$\Phi^{RA}(x_1, x_2, x_3, x_4)$$
$$= \int D\boldsymbol{\Psi} \varphi_\alpha(x_1) \varphi_\alpha(x_2) \phi_\beta(x_3) \phi_\beta(x_4)$$
$$\times \int_{-\infty}^{\infty} d\xi e^{i\Delta\xi/2} \delta(\xi - \int d^d x(\boldsymbol{\varphi}^2 - \boldsymbol{\phi}^2)) \quad (66)$$
$$\times \exp\left\{-\int d^d x \left[\frac{1}{2}(\nabla \boldsymbol{\Psi})^2 + \frac{1}{2}\kappa^2 \boldsymbol{\Psi}^2 + \frac{1}{4}g\boldsymbol{\Psi}^4\right]\right\},$$

where

$$\vec{\boldsymbol{\Psi}} = \begin{Bmatrix} \boldsymbol{\varphi} \\ \boldsymbol{\phi} \end{Bmatrix}. \quad (67)$$

is a composite $(m + n)$-component vector. The saddle-point configuration in instanton calculations is sought in the form

$$\boldsymbol{\varphi}_c(x) = \frac{\psi_c(x)}{\sqrt{-g}} \mathbf{u}, \quad \boldsymbol{\phi}_c(x) = \frac{\psi_c(x)}{\sqrt{-g}} \mathbf{v},$$
$$\boldsymbol{\Psi}_c(x) = \frac{\psi_c(x)}{\sqrt{-g}} \mathbf{U}, \quad (68)$$

where

$$\mathbf{U} = \begin{Bmatrix} \mathbf{u} \\ \mathbf{v} \end{Bmatrix}, \quad U^2 = u^2 + v^2. \quad (69)$$

It is known that the expansion coefficients for the correlation function of $M$ fields in the $n$-component $\varphi^4$ theory have the form

$$\langle \varphi_{\alpha_1}(x_1) \ldots \varphi_{\alpha_M}(x_M) \rangle_N = c\{x_i\}(-S_0)^{-N}$$
$$\times \Gamma(N + b) \int d^n u \delta(u^2 - 1) u_{\alpha_1} \ldots u_{\alpha_M},$$
$$c\{x_i\} = C_{Mn} \int d^d y \psi_c(x_1 - y) \ldots \psi_c(x_M - y), \quad (70)$$
$$S_0 = \frac{I_4}{4}, \quad I_p = \int d^d x \psi_c^p(x),$$

$$b = \begin{cases} (M + n + d - 1)/2, & d < 4, \\ (M + n - 1)/2, & d > 4. \end{cases}$$

The expression for the constant $C_{Mn}$ is lengthy and can be found in [13, 27]. If the saddle-point approximation

$$\delta\left(\xi - \frac{I_2}{|g_c|(u^2 - v^2)}\right), \quad g_c = -\frac{I_4}{4N}, \quad (71)$$

is used for the $\delta$ function term entering into Eq. (66), the integral with respect to $D\Psi$ in Eq. (66) is a usual functional integral of the $(m + n)$-component $\varphi^4$ theory and its expansion coefficients follow from Eq. (70) at $M = 4$ as a result of change of the integration with respect to $d^n u$ to the integration with respect to $d^{n+m}U = d^n u d^m v$:

$$[\Phi^{RA}\{x_i\}]_N = c\{x_i\}(-S_0)^{-N}\Gamma(N + b) \\ \times \int d^n u d^m v \delta(u^2 + v^2 - 1) u_\alpha^2 v_\beta^2 e^{i\Delta\lambda N(u^2 - v^2)}, \quad (72)$$

where $\lambda = 2I_2/I_4$. The calculation of the integral is performed in the coordinates $u = r\cos\theta$ and $v = r\sin\theta$ and, in the limit $n, m \longrightarrow 0$, provides

$$\frac{S_n}{n}\frac{S_m}{m}\int_0^\infty u^{n-1}du\int_0^\infty v^{m-1}dv\delta(u^2 + v^2 - 1)u^2 v^2 \\ \times e^{i\Omega(u^2 - v^2)} = \frac{e^{i\Omega} - e^{-i\Omega}}{8i\Omega}. \quad (73)$$

Here, the following expression is used:

$$\frac{S_n}{n} = \frac{2\pi^{n/2}}{n\Gamma(n/2)} \longrightarrow 1 \quad (n \longrightarrow 0). \quad (74)$$

where $S_n$ is the area of the $n$-dimensional unit sphere. Thus, the high-order asymptotic expression for the coefficients of the expansion of $\Phi^{RA}$,

$$[\Phi^{RA}\{x_i\}]_N = \frac{1}{4}c\{x_i\}(-S_0)^{-N}\Gamma(N + b) \\ \times \frac{e^{i\Delta\lambda N} - e^{-i\Delta\lambda N}}{2i\Delta\lambda N} \quad (75)$$

has the same $\Delta$ dependence as in the zero-dimensional case [see Eq. (50)]. For this reason, the argumentation given in Section 6 for the necessity of the imaginary additions of opposite signs to $\kappa_1^2$ and $\kappa_2^2$ for the appearance of a singular contribution in the summation of high orders,

$$[\Phi^{RA}\{x_i\}]_{\text{sing}} = \frac{I_4}{8I_2}c\{x_i\}\left(\frac{S_0}{|g|}\right)^b \\ \times \exp\left\{-\frac{S_0}{|g|}\right\}\frac{4\pi}{-i\omega + 2\gamma}. \quad (76)$$

remains valid. According to the above analysis, the infinitely small imaginary addition $\delta$ appearing in Eq. (9) is changed to the finite value $\gamma$, which remains indefinite in the framework of the instanton analysis.

In the above analysis, it is implicitly implied that the dimension of space is $d < 4$ and the instanton $\psi_c(x)$ is determined from the equation

$$-\nabla\psi_c + \kappa^2 - \psi_c^3 = 0. \quad (77)$$

The theory for $d > 4$ is nonrenormalizable and should be considered on a lattice. This requires the modification of the gradient term in the action and the change of the integration with respect to $d^d y$ in Eq. (70) to the summation over the sites of the lattice. Results (75) and (76) remain unchanged, but $\psi_c(x)$ should be treated as a lattice instanton [13]. The above calculations at $d = 4$ should be modified: the integral $I_2$ diverges in massless theory and, in theory with mass, requires accurate integration over an additional soft mode associated with change in the radius of the instanton [26].

ACKNOWLEDGMENTS

This work was supported by the Russian Foundation for Basic Research (project no. 06-02-17541).